\documentclass[epsfig,usenatbib]{mnras}
\usepackage{epsfig}
\usepackage{natbib}
\usepackage{amsmath}

\def\gtsima{$\; \buildrel > \over \sim \;$}
\def\ltsima{$\; \buildrel < \over \sim \;$}
\def\gtrsim{\lower.5ex\hbox{\gtsima}}
\def\lesssim{\lower.5ex\hbox{\ltsima}}

\begin{document}
\title[The cosmic merger rate of NSs and BHs]{The cosmic merger rate of neutron stars and black holes}
\author[Mapelli \& Giacobbo]
{Michela Mapelli$^{1,2,3}$ \&{} Nicola Giacobbo$^{1,2,3,4}$ 
\\
$^1$Institut f\"ur Astro- und Teilchenphysik, Universit\"at Innsbruck, Technikerstrasse 25/8, A--6020, Innsbruck, Austria\\
$^2$INAF-Osservatorio Astronomico di Padova, Vicolo dell'Osservatorio 5, I--35122, Padova, Italy, {\tt michela.mapelli@oapd.inaf.it}\\
$^3$INFN, Milano Bicocca, Piazza della Scienza 3, I--20126 Milano, Italy\\
$^4$Physics and Astronomy Department Galileo Galilei, University of Padova, Vicolo dell'Osservatorio 3, I--35122, Padova, Italy\\
}
\maketitle \vspace {7cm }
\bibliographystyle{mnras}
 
\begin{abstract}
Six gravitational wave detections have been reported so far, providing crucial insights on the merger rate of double compact objects. We investigate the cosmic merger rate of double neutron stars (DNSs), neutron star-black hole binaries (NSBHs) and black hole binaries (BHBs) by means of population-synthesis simulations coupled with the Illustris cosmological simulation. We have performed six different simulations, considering different assumptions for the efficiency of common envelope (CE) ejection and exploring two distributions for the supernova (SN) kicks. The current BHB merger rate derived from our simulations spans from $\sim{}150$ to $\sim{}240$ Gpc$^{-3}$ yr$^{-1}$ and is only mildly dependent on CE efficiency. In contrast, the current merger rates of DNSs (ranging from $\sim{}20$ to $\sim{}600$ Gpc$^{-3}$ yr$^{-1}$) and NSBHs (ranging from $\sim{}10$ to $\sim{}100$ Gpc$^{-3}$ yr$^{-1}$) strongly depend on the assumptions on CE and natal kicks. The merger rate of DNSs is consistent with the one inferred from the detection of GW170817 only if a high efficiency of CE ejection and low SN kicks (drawn from a Maxwellian distribution with one dimensional root mean square $\sigma{}=15$ km s$^{-1}$) are assumed.
\end{abstract}

%

\section{Introduction}
The LIGO-Virgo collaboration \citep{LIGOdetector,VIRGOdetector} has reported six gravitational wave (GW) events so far, five of them involving the merger of two black holes (BHs, \citealt{abbott2016a,abbott2016b,abbott2016c,abbott2016d,abbott2017a,abbott2017b,abbott2017c}) and one (GW170817) interpreted as the merger of two neutron stars (NSs, \citealt{abbott2017d,abbott2017e}). The merger rate of NSs and BHs can be inferred from these six detections, providing some crucial insights on the formation channels of compact-object binaries.


Recent estimates of the BH binary (BHB) merger rate from GW detections span from $\sim{}9$ to $\sim{}240$  events Gpc$^{-3}$ yr$^{-1}$ \citep{abbott2016d}, while for NS--BH binaries (NSBHs) an upper limit of $\sim{}3600$ events Gpc$^{-3}$ yr$^{-1}$ \citep{abbott2016e} was derived from the first observing (O1) run. Finally, a rate of $1540_{-1220}^{+3200}$ Gpc$^{-3}$ yr$^{-1}$ was inferred for double NS (DNS) mergers, based on GW170817 \citep{abbott2017d}.

Alternative observational estimates of the DNS merger rate are based either on the rate of short gamma-ray bursts \citep{coward2012,petrillo2013,siellez2014,fong2015}, or on kilonova observations \citep{jin2015}, or on the properties of Galactic DNSs \citep{kim2015}. Overall, the DNS merger rate derived from gamma-ray bursts and from kilonovae (ranging from few tens to several thousands of mergers Gpc$^{-3}$ yr$^{-1}$) is consistent with the one inferred from GW events. 

From a theoretical perspective, predicting the cosmic merger rate of compact objects is quite an ordeal, because of the involved dynamical range: we need to put the mergers of compact-object binaries (with an initial orbital separation of tens of solar radii) in a large cosmological framework. The early work of \cite{dominik2013} achieves this result by combining the results of their population-synthesis simulations with  a description of the cosmic star-formation rate (SFR) density and of the average metallicity evolution, predicting a local merger rate of  $\sim{}30$, $\sim{}60$ and $\sim{}2$ Gpc$^{-3}$ yr$^{-1}$ for BHB, DNS and NSBH mergers, respectively, according to their fiducial model. Since then, many studies \citep{belczynski2016,dvorkin2016,elbert2017,lamberts2016,oshaughnessy2017,lamberts2018,cao2018} have focused on the cosmic merger rate of BHBs adopting a number of approaches. In particular, \cite{schneider2017} and \cite{mapelli2017} have combined catalogues of merging BHBs from population-synthesis simulations with the outputs of cosmological simulations. Following this approach, \cite{mapelli2017} find a local BHB merger rate of $\sim{}120-180$ Gpc$^{-3}$ yr$^{-1}$ for their fiducial model, consistent with the results of the LIGO-Virgo collaboration.

As to DNS mergers, population-synthesis simulations have been used to predict merging rates spanning from few tens to several hundreds mergers Gpc$^{-3}$ yr$^{-1}$ (e.g. \citealt{abadie2010,dominik2013,dominik2015,ziosi2014,deminkbelczynski2015}). Recently, \cite{chruslinska2018} have claimed that their population-synthesis models can obtain a local DNS merger rate consistent with GW detections only if they use a very optimistic description for the common-envelope (CE) phase, assuming that Hertzsprung gap (HG) donor stars can avoid merger during CE. This description of CE, if applied to merging BHBs, overestimates the merger rate of BHBs by a large factor. Their main conclusion is that either CE works differently for DNSs and BHBs, or high natal kicks for BHs are needed to match the results of GW detections. Similarly, \cite{kruckow2018} obtain a local DNS merger rate density of only $\sim{}10$ Gpc$^{-3}$ yr$^{-1}$ for their fiducial model and a strong upper limit to the local DNS merger rate of $\sim{}400$ Gpc$^{-3}$ yr$^{-1}$.

In this paper, we use the same approach as \cite{mapelli2017} to estimate the cosmic merger rate of BHBs, DNSs and NSBH binaries: we use a Monte Carlo code to combine the results of advanced population-synthesis simulations \citep{giacobbo2018a,giacobbo2018b} with the outputs of the Illustris-1 cosmological simulation \citep{vogelsberger2014a}. This method enables us to account for the cosmic SFR density and for the metallicity evolution in the host galaxies, within the limitations posed by state-of-the-art cosmological simulations.

\section{Methods}\label{sec:methods}
\subsection{Population-synthesis simulations with \sc MOBSE}
We have run a large grid of binary population-synthesis simulations with the code {\sc MOBSE} \citep{giacobbo2018a}, to derive catalogues of merging DNSs, NSBHs and BHBs. {\sc MOBSE} is an upgraded version of {\sc BSE} \citep{hurley2000,hurley2002}. The main updates of {\sc MOBSE} with respect to {\sc BSE} are the following (see \citealt{giacobbo2018a} and \citealt{giacobbo2018b} for details).
\begin{itemize}
\item[-]Stellar winds of hot massive stars (O- and B- type, Wolf-Rayet and luminous blue variable stars) have been updated including a dependence of mass loss on metallicity as $\dot{M}\propto{}Z^{\beta{}}$, with $\beta=0.85$ if $\Gamma{}_e<2/3$,  $\beta=2.45-2.4\,{}\Gamma_e$ if $2/3\leq{}\Gamma{}_e\le{}1$, and $\beta{}=0.05$ if $\Gamma{}_e\ge{}1$ (where $\Gamma{}_e$ is the electron-scattering Eddington ratio, i.e. the ratio between current stellar luminosity and Eddington luminosity, \citealt{vink2001,vink2005,graefener2008,chen2015}).
\item[-]The core radii of massive evolved stars have been updated as in \cite{halltout2014}.
\item[-]Core-collapse supernovae (SNe) are described through the rapid and the delayed SN mechanism described in \cite{fryer2012}. In the simulations performed for this paper we adopt the \emph{rapid} core-collapse SN model (changing the core-collapse SN model from rapid to delayed would have a minor effect on the merger rate, as already shown in Figure~1 of \citealt{mapelli2017}).
\item[-]Pulsational pair-instability and pair-instability SNe are included, according to \cite{spera2017} (see also \citealt{woosley2017} and \citealt{giacobbo2018a}).
\item[-]Electron-capture SNe (ECSNe) are described as in \cite{hurley2000}, the only update being the minimum Oxygen-Neon core mass for an ECSN to occur (which is set to be 1.38 M$_\odot$ instead of 1.44 M$_\odot$, see \citealt{giacobbo2018c}).    
\item[-]The natal kick of NSs is obtained by randomly picking a number according to a Maxwellian distribution. For ECSNe, the Maxwellian distribution is assumed to have a one dimensional root mean square (1D rms) $\sigma_{\rm ECSN}=15$ km s$^{-1}$, which was found to maximize the number of merging DNSs in \cite{giacobbo2018c}. For core-collapse SNe, the Maxwellian distribution has a 1D rms $\sigma_{\rm CCSN}=265$ km s$^{-1}$ (consistent with the proper motions of 233 single pulsars reported in \citealt{hobbs2005}) in the fiducial case, and a 1D rms  $\sigma_{\rm CCSN}=15$ km s$^{-1}$ in the runs labeled as ``low $\sigma{}$''  (see Table~\ref{tab:table1}).

  For BHs, the amount of kick is scaled by the fallback as $v_{\rm BH}=v_{\rm NS}\,{}(1-f_{\rm fb})$, where $v_{\rm BH}$ is the natal kick for the BH, $v_{\rm NS}$ is the natal kick for a NS (drawn from a Maxwellian distribution  as described above), and $f_{\rm fb}$ (ranging from 0 to 1) is the amount of  fallback on the proto-NS \citep{fryer2012,spera2015}.
\item[-]We describe common envelope (CE) with the same formalism as in \cite{hurley2002}. This formalism depends on two free parameters, $\alpha{}$ (describing the efficiency of energy transfer) and $\lambda{}$ (describing the geometry of the envelope and the importance of recombinations). In the current paper, $\lambda{}$ is defined as in \cite{claeys2014}, to account for the contribution of recombinations, while $\alpha$ is a constant.

  The main change in the description of CE with respect to {\sc BSE} consists in the treatment of Hertzsprung gap (HG) stars. In the standard version of \textsc{BSE},  HG donors entering a CE phase are allowed to survive the CE phase. In \textsc{MOBSE}, HG donors are forced to merge with their companion if they enter a CE. Models in which HG donors are allowed to survive a CE phase produce a local BHB merger rate $R_{\rm BHB}\sim{}600-800$ Gpc$^{-3}$ yr$^{-1}$, which is not consistent with LIGO-Virgo results \citep{mapelli2017}.
\end{itemize}


We have run six different sets of population-synthesis simulations with this version of {\sc MOBSE}, adopting $\alpha{}=1,3,$ and 5 and by changing the value of $\sigma_{\rm CCSN}=15,\,{}265$ km s$^{-1}$. The details of the six runs can be found in Table~\ref{tab:table1}.

For each of the  six simulation sets described in Table~\ref{tab:table1}, we have simulated  12 sub-sets with metallicity $Z=0.0002,$ 0.0004, 0.0008, 0.0012, 0.0016, 0.002, 0.004, 0.006, 0.008, 0.012, 0.016, and 0.02.   Throughout the paper, we define solar metallicity as Z$_\odot{}=0.02$. Thus, the 12 sub-sets correspond to metallicity $Z=0.01$,   0.02,   0.04,   0.06,   0.08,   0.1,   0.2,   0.3,   0.4,   0.6,   0.8, and 1.0~Z$_\odot$. In each sub-set  we  have simulated $10^7$ stellar binaries. Thus, each of the  six sets of simulations is composed of $1.2\times{}10^8$ massive binaries.

For each binary, the mass of the primary ($m_{\rm p}$) is randomly drawn from a Kroupa initial mass function \citep{kroupa2001} ranging\footnote{The fitting formulas by \cite{hurley2000} might be inaccurate for very massive stars. To improve the treatment of massive stars, we impose that the values of the radius of single stars are consistent with \textsc{PARSEC} stellar evolution tracks \citep{chen2015} for stars with mass $>100$ $M_\odot$, as discussed in \cite{mapelli2016}.} from 5 to 150 M$_\odot$, and the mass of the secondary ($m_{\rm s}$) is sampled according to the distribution $\mathcal{F}(q)\propto{}q^{-0.1}$ (where $q=m_{\rm s}/m_{\rm p}$) in a range  $[0.1\,{}-1]\,{}m_{\rm p}$. The orbital period $P$ and the eccentricity $e$ are randomly extracted from the distribution $\mathcal{F}(P)\propto{}(\log_{10}{P})^{-0.55}$, with $0.15\leq{}\log_{10}{(P/{\rm day})}\leq{}5.5$, and $\mathcal{F}(e)\propto{}e^{-0.42}$, with $0\leq{}e<1$, as suggested by \cite{sana2012}.

\begin{table}
\begin{center}
\caption{\label{tab:table1}
Properties of the population-synthesis simulations.}
 \leavevmode
\begin{tabular}[!h]{lll}
\hline
Name    & $\alpha{}$ & $\sigma_{\rm CCSN}$\\
& & [km s$^{-1}$]\\
\hline
$\alpha=1$  & 1.0        & 265  \\
$\alpha=1$, low $\sigma{}$  & 1.0        & 15  \\
\noalign{\vspace{0.1cm}}
$\alpha=3$  & 3.0        & 265  \\
$\alpha=3$, low $\sigma{}$  & 3.0        & 15  \\
\noalign{\vspace{0.1cm}}
$\alpha=5$  & 5.0        & 265  \\
$\alpha=5$, low $\sigma{}$  & 5.0        & 15  \\
\noalign{\vspace{0.1cm}}
\hline
\end{tabular}
\begin{flushleft}
\footnotesize{Column 1: model name; column 2: value of $\alpha{}$ in the CE formalism; column 3: 1D rms of the Maxwellian distribution for the core-collapse SN kick. See the text for details. }
\end{flushleft}
\end{center}
\end{table}
\subsection{The Illustris cosmological simulation}\label{sec:section2.2}
The Illustris-1 is the highest resolution hydrodynamical simulation run in the frame of the Illustris project \citep{vogelsberger2014a,vogelsberger2014b,nelson2015}. In the following, we refer to it simply as the Illustris. It covers a comoving volume of $(106.5\,{}{\rm Mpc})^3$, and has an initial dark matter and baryonic matter mass resolution of $6.26\times{}10^6$ and $1.26\times{}10^6$ M$_\odot$, respectively \citep{vogelsberger2014a,vogelsberger2014b}. At redshift zero the softening length is $\sim{}710$ pc, while the  smallest hydrodynamical cells have a length of 48 pc.

The size of the Illustris' box ensures that we include the most massive haloes. 
 On the other hand, the population of dwarf galaxies is under-resolved. In \cite{mapelli2017} we estimate in less than $\sim{}40$ per cent the effect of resolution limitations on the merger rate. In a forthcoming paper, we will describe the contribution of dwarf galaxies to the merger rate by adopting a smaller box, higher resolution cosmological simulation. At redshift $z <0.3$ the Illustris simulation predicts a higher SFR density by a factor of $\sim{}1.5$ with respect to the observed one \citep{madaudickinson2014}. This suggests that the local merger rate density might be slightly overestimated in our calculations.

The Illustris includes a treatment for sub-grid physics (cooling, star formation, SNe, super-massive BH formation, accretion and merger, AGN feedback, etc), as described in \cite{vogelsberger2013}. The model of sub-grid physics adopted in the Illustris is known to produce a mass-metallicity relation \citep{genel2014,genel2016} which is sensibly steeper than the observed one (see the discussion in \citealt{vogelsberger2013} and \citealt{torrey2014}). Moreover, the simulated mass-metallicity relation does not show the observed turnover at high stellar mass ($\gtrsim{}10^{10}$ M$_\odot{}$). In \cite{mapelli2017}, we estimate that the impact on the BHB merger rate of the difference between the Illustris mass-metallicity relation and the observational relation \citep{maiolino2008,mannucci2009} is of the order of 20 per cent. As shown in \cite{giacobbo2018b}, the merger rate of DNSs does not significantly depend on the metallicity of their progenitors. Thus, the DNS merger rate is essentially unaffected by the deviation between the mass-metallicity relation of the Illustris and the observed one.

As for the cosmology, the Illustris adopts WMAP-9 results for the cosmological parameters \citep{hinshaw2013}, that is $\Omega{}_{\rm M} = 0.2726$, $\Omega{}_\Lambda = 0.7274$, $\Omega{}_b = 0.0456$, and $H_0 = 100\,{}h$ km s$^{-1}$ Mpc$^{-1}$, with $h = 0.704$. 

\subsection{Planting BHBs into a cosmological simulation}
We combine the catalogues of merging DNSs, NSBHs and BHBs with the Illustris simulations as described in \cite{mapelli2017}. In particular, we store the initial mass $M_{\rm Ill}$, formation redshift $z_{\rm Ill}$ and metallicity $Z_{\rm Ill}$ of each stellar particle of the Illustris. From the {\sc MOBSE} simulations, we extract the compact-object masses and the delay time $t_{\rm delay}$ (i.e. the time elapsed between the formation of the progenitor stars and the merger of the compact-object binary) for all compact-object binaries merging in  less than  a Hubble time. We also extract the total initial mass $M_{\rm BSE}$ of the stellar population simulated with {\sc MOBSE} for a given run and metallicity.

Through a Monte Carlo scheme, we associate to each newly born Illustris stellar particle (see equation~1 of \citealt{mapelli2017}) a number $n_{\rm CO,\,{}i}$ of compact-object binaries (where the index $i=$ BHB, NSBH or DNS) as given by 
\begin{equation}
n_{\rm CO,\,{}i}=N_{\rm BSE,\,{}i}\,{}\frac{M_{\rm Ill}}{M_{\rm BSE}}\,{}f_{\rm corr}\,{}f_{\rm bin},
\end{equation}
where  $N_{\rm BSE,\,{}i}$ is the number of merging compact objects within the simulated sub-set of initial stellar mass $M_{\rm BSE}$ (with $i=$ BHB, NSBH or DNS), 
and $f_{\rm corr}= 0.285$ is a correction factor, accounting for the fact that we actually simulate only primaries with $m_{\rm p}\ge{}5$ M$_\odot$, neglecting lower mass stars. Finally, $f_{\rm bin}$ accounts for the fact that we simulate only binary systems, whereas a fraction of stars are single. Here we assume that 50 per cent of stars are in binaries, thus $f_{\rm bin}=0.5$. We note that  $f_{\rm bin}$ is only a scale factor and our results can be re-scaled to a different  $f_{\rm bin}$  a posteriori. 

We associate an Illustris stellar particle to a population-synthesis simulation set by looking for the population-synthesis simulation set with the closest metallicity to $Z_{\rm Ill}$  among the 12 metallicities simulated with \textsc{MOBSE}\footnote{If  $Z_{\rm Ill}>0.02$ ($Z_{\rm Ill}<0.0002$), we associate to the Illustris's particle a \textsc{MOBSE} set with  $Z_{\rm Ill}=0.02$ ($Z_{\rm Ill}=0.0002$), since  the maximum (minimum) metallicity we simulated with \textsc{BSE} is 0.02 (0.0002).}.

We then estimate the look-back time of the merger ($t_{\rm merg}$) of each compact-object binary in the randomly selected sample as $t_{\rm merg}=t_{\rm form}-t_{\rm delay}$, where $t_{\rm delay}$ is the time between the formation of the progenitor stellar binary and the merger of the compact-object binary, and $t_{\rm form}$ is the look back time at which the Illustris' particle has formed, calculated as
\begin{equation}
t_{\rm form}=\frac{1}{H_0}\int_0^{z_{\rm Ill}}\frac{1}{(1+z)\,{}\left[\Omega{}_{\rm M}\,{}(1+z)^3+\Omega{}_\Lambda{}\right]^{1/2}}{\rm d}z,
\end{equation}
where the cosmological parameters are set to WMAP-9 values (for consistency with the Illustris) and $z_{\rm Ill}$ is the formation redshift of the Illustris' particle.

According to this definition, $t_{\rm merg}$ is also a look back time: it tells us how far away from us the compact objects have merged. For our analysis, we consider only compact objects with $t_{\rm merg}\ge{}0$, i.e. we do not consider compact objects that will merge in the future. We repeat the same procedure for each of the  six simulation sets in Table~\ref{tab:table1} and we obtain six different models of the cosmic merger evolution for BHBs, NSBHs and DNSs.

\section{Results}\label{sec:results} 

\begin{figure}
\center{{
\epsfig{figure=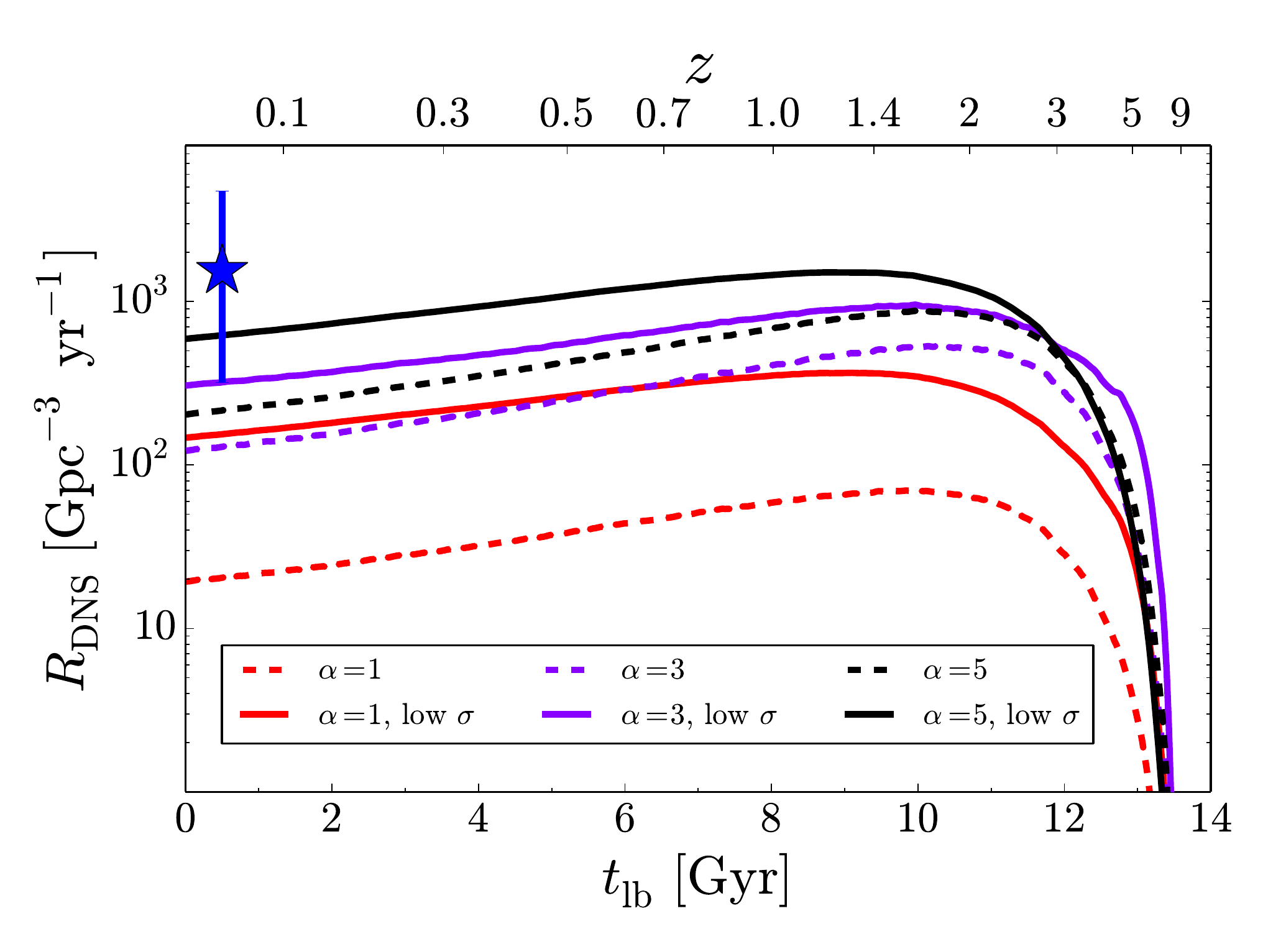,width=9cm} 
}}
\caption{\label{fig:figDNS}
  Cosmic merger rate density of DNSs ($R_{\rm DNS}$) in the comoving frame, as a function of the look-back time $t_{\rm lb}$ (bottom $x$ axis) and of the redshift $z$ (top $x$ axis) in our models (see Table~\ref{tab:table1}).
  Blue error bar with star: DNS merger rate inferred from GW170817 \citep{abbott2017d}. The position of the data point along the $x$ axis (corresponding to $t_{\rm lb}=0.5$ Gyr) is just for visualization purposes. 
}
\end{figure}

\begin{figure}
\center{{
\epsfig{figure=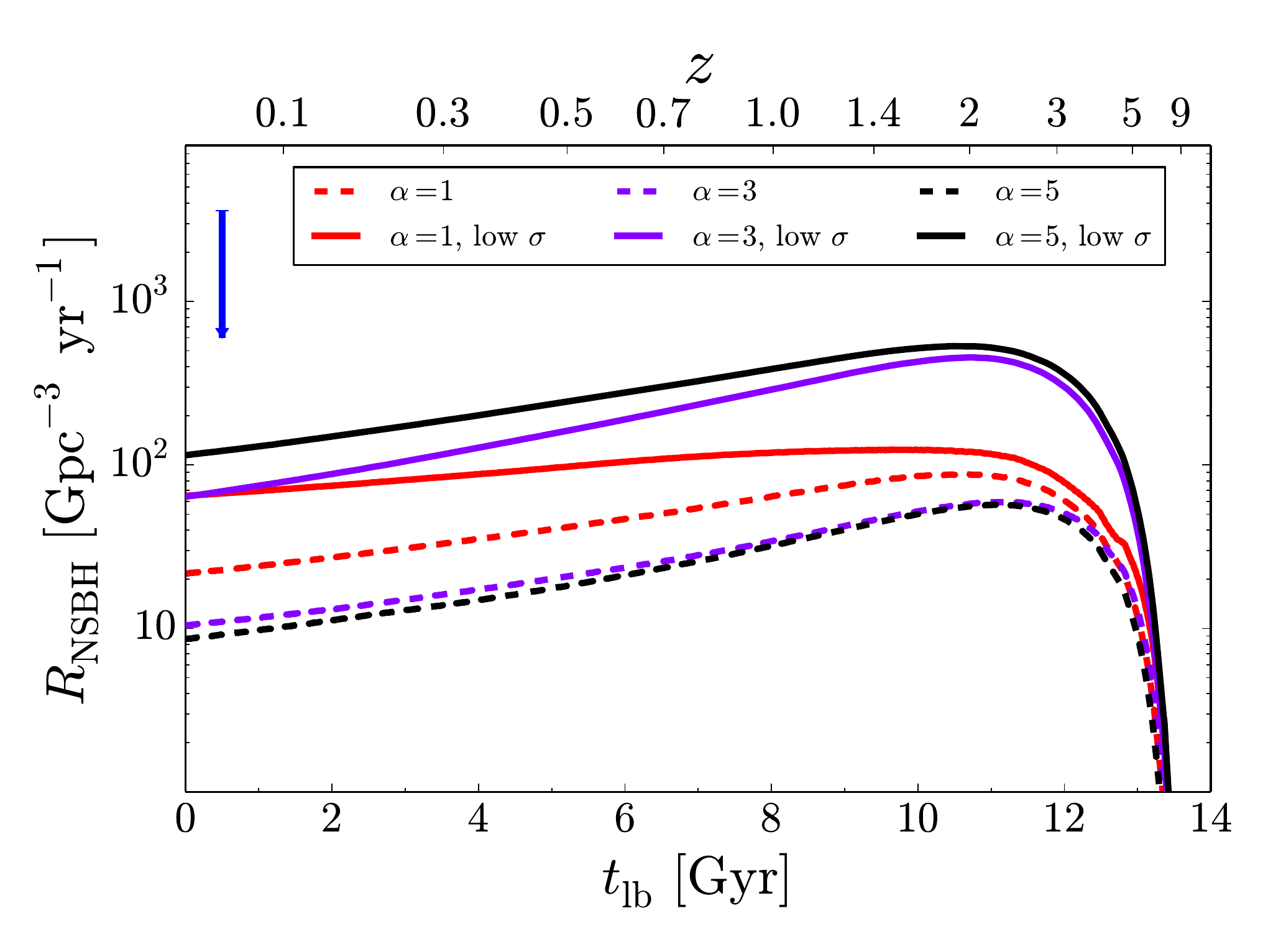,width=9cm} 
}}
\caption{\label{fig:figNSBH}
  Cosmic merger rate density of NSBHs ($R_{\rm NSBH}$) in the comoving frame, as a function of the look-back time $t_{\rm lb}$ (bottom $x$ axis) and of the redshift $z$ (top $x$ axis) in our models  (see Table~\ref{tab:table1}).
  Blue arrow: upper limit to the NSBH merger rate inferred from the LIGO O1 run \citep{abbott2016e}. The position of the upper limit along the $x$ axis (corresponding to $t_{\rm lb}=0.5$ Gyr) is just for visualization purposes. 
}
\end{figure}

\begin{figure}
\center{{
\epsfig{figure=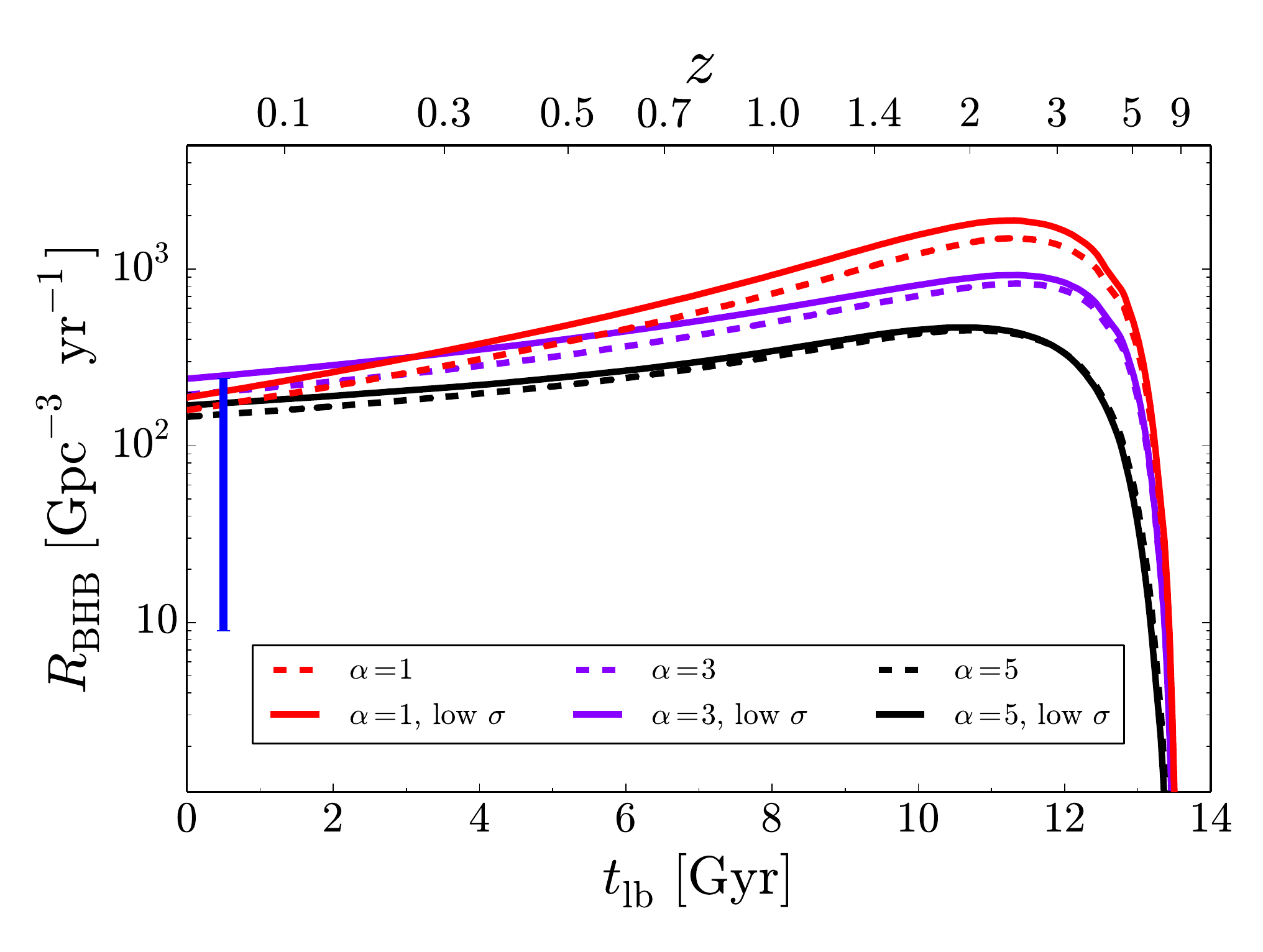,width=9cm} 
}}
\caption{\label{fig:figBHB}
  Cosmic merger rate density of BHBs ($R_{\rm BHB}$) in the comoving frame, as a function of the look-back time $t_{\rm lb}$ (bottom $x$ axis) and of the redshift $z$ (top $x$ axis) in our models  (see Table~\ref{tab:table1}).
  Blue error bar: BHB merger rate inferred from the LIGO O1 run \citep{abbott2016d}. The position of the data point along the $x$ axis (corresponding to $t_{\rm lb}=0.5$ Gyr) is just for visualization purposes. 
}
\end{figure}

\subsection{Merger rate}\label{sec:mergrate}

 The compact-object binary  merger rate density in the comoving frame $R_{\rm CO,\,{}i}$ (with $i=$ BHB, NSBH or DNS) can be derived from our simulations as:
\begin{equation}\label{eq:DNSrate}
R_{\rm CO,\,{}i}=N_{\rm CO,\,{}i}\,{}{\rm Gpc}^{-3}\,{}{\rm yr}^{-1}\,{}\left(\frac{l_{\rm box}}{{\rm Gpc}}\right)^{-3}\,{}\,{}\left(\frac{\Delta{}t}{\rm yr}\right)^{-1},
\end{equation}
where $N_{\rm CO,\,{}i}$ is the number of compact-object mergers (with $i=$ BHB, NSBH or DNS) per time bin $\Delta{}t$ in the entire Illustris box, $l_{\rm box}=106.5$ Mpc is the size of the Illustris box (in the comoving frame) and $\Delta{}t$ is the size of the time bin (we adopt $\Delta{}t=$10 Myr).

The resulting merger rate $R_{CO,\,{}i}$ is affected by stochastic fluctuations due to the Monte Carlo sampling. By running several random realisations of the population synthesis simulations, we have estimated the fluctuations of $R_{\rm CO,\,{}i}$ to be less than few per cent between two different realisations.

Figure~\ref{fig:figDNS} shows the cosmic merger rate density of DNSs ($R_{\rm DNS}$) in the comoving frame, derived from our simulations  using equation~\ref{eq:DNSrate}. The corresponding values of the merger rate at redshift $z=0$ are shown in Table~\ref{tab:table2}.

From Fig.~\ref{fig:figDNS} it is apparent that our estimate of the DNS merger rate is consistent with the rate inferred from GW170817 \citep{abbott2017d} only if we assume a large value of the CE parameter $\alpha{}$ ($\alpha{}\ge{}3$) and low kicks for both ECSNe and core-collapse SNe. Lower values of $\alpha$ and larger kicks produce lower values of the DNS merger rate.

The magnitude of the natal kick is particularly important for DNSs, because a large kick can easily unbind the binary. A large value of $\alpha{}$ means that CE ejection is particularly efficient, preventing a binary to merge after CE. 

The DNS merger rate increases with redshift, reaching a maximum at $z\sim{}1.5-2.5$, depending on the model. This result is consistent with the trend found by \cite{mapelli2017} for BHBs and is the effect of a convolution between the cosmic SFR density and the distribution of delay times.

Figure~\ref{fig:figNSBH} shows our prediction for the NSBH merger rate in the comoving frame ($R_{\rm NSBH}$), obtained from equation~\ref{eq:DNSrate} when considering the number of NSBH mergers per time bin ($N_{\rm NSBH}$). The estimated merger rate of NSBHs is consistent with  the upper limit from O1 in all considered models. The merger rate of NSBHs increases with redshift, reaching a maximum at $z\sim{}1.5-3$, similar to the behaviour of DNSs.

Models with low kicks produce higher local NSBH merger rates by a factor of $3-10$ with respect to models with large SN kicks. If low kicks are assumed for the core-collapse SNe, the NSBH merger rate increases by increasing $\alpha{}$. In contrast, if larger kicks are assumed for core-collapse SNe, a larger value of $\alpha{}$ reduces the number of NSBH mergers.
 
This happens because a large core-collapse SN kick increases the semi-major axis of the NSBH or even breaks the system. Binary systems that went through a CE phase with a small value of $\alpha{}$ are more likely to avoid being broken and to remain sufficiently bound after the second core-collapse SN, because a small value of $\alpha{}$ implies that the system's semi-major axis shrinks considerably during CE, before the CE is ejected. In contrast, a large value of $\alpha{}$ means that the CE is expelled without much shrinking of the orbital separation and makes easier for the second SN to unbind the system.


Finally, Fig.~\ref{fig:figBHB} shows the merger rate of BHBs,  calculated adopting the number of BHB mergers per time bin ($N_{\rm BHB}$)  in equation~\ref{eq:DNSrate}. The estimated BHB merger rate is consistent with the range inferred from the O1 run, although close to the upper limit (90\% credible level). In contrast with DNSs, different values of $\alpha{}$ and of the SN kick do not seem to affect the BHB merger rate significantly.

This happens because the natal kicks of BHs are quite low  in our models, even if $\sigma{}_{\rm CCSN}=265$ km s$^{-1}$ is assumed, because the kick is modulated by fallback. The effect would have been much more important if we would not have assumed a dependence of the kick on the amount of fallback, as shown by model DK of \cite{mapelli2017} (see their Figure~1).

The BHB merger rate increases with redshift, reaching a maximum at $z\sim{}2-4$. The increase of the merger rate with redshift is much steeper for BHBs than for DNSs and NSBHs, because of the dependence of BHB mergers on metallicity. As shown in Figure~12 of \cite{giacobbo2018a}, the number of BHB mergers per unit mass is $\sim{}2$ orders of magnitude larger at low metallicity ($Z\leq{}0.002$) than at high metallicity. This means that we expect a much higher BHB merger rate in the early Universe (which is predominantly metal poor) with respect to the nearby Universe (where high metallicity is more common). In contrast, the merger rate of DNSs does not seem to depend on metallicity.

\begin{table}
\begin{center}
\caption{\label{tab:table2}
Comoving merger-rate density of DNSs NSBHs and BHBs at redshift $z=0$ ($t_{\rm lb}=0$).} 
 \leavevmode
\begin{tabular}[!h]{llll}
\hline
Name &  $R_{\rm DNS}$ & $R_{\rm NSBH}$ & $R_{\rm BHB}$\\ 
     & [Gpc$^{-3}$ yr$^{-1}$] & [Gpc$^{-3}$ yr$^{-1}$] & [Gpc$^{-3}$ yr$^{-1}$]\\
\hline
$\alpha=1$  & 19 & 22 & 159 \\
$\alpha=1$, low $\sigma{}$  & 147 & 65 & 188\\
$\alpha=3$  & 122 & 10 & 195\\
$\alpha=3$, low $\sigma{}$  & 306 & 64 & 240 \\
$\alpha=5$  & 204 & 9 & 146 \\
$\alpha=5$, low $\sigma{}$  & 591 & 115 & 169\\
\noalign{\vspace{0.1cm}}
\hline
\end{tabular}
\begin{flushleft}
\footnotesize{Column 1: model name; column 2: DNS merger rate density at $z=0$;  column 3: NSBH merger rate density at $z=0$; column 4: BHB merger rate density at $z=0$.  Uncertainties in the tabulated rates are of the order of few per cent, because of stochastic fluctuations.}
\end{flushleft}
\end{center}
\end{table}
\subsection{Mass distribution}\label{sec:massdistr}
\begin{figure}
\center{{
\epsfig{figure=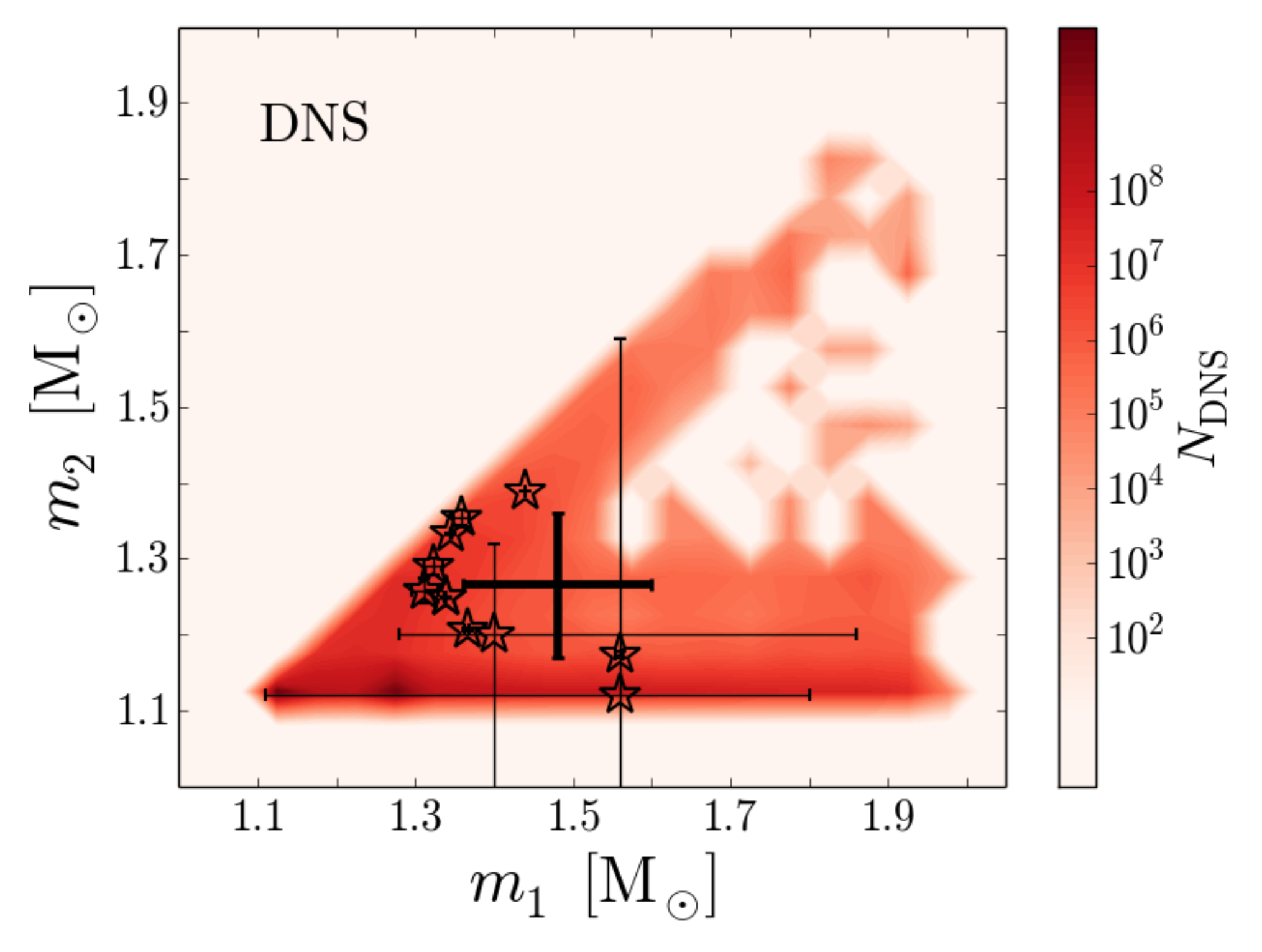,width=9cm} 
}}
\caption{\label{fig:figNSNSmass}
  Mass of the primary NS versus the mass of the secondary NS for all simulated merging DNSs for the run with $\alpha=5$ and $\sigma{}_{\rm CCSN}=15$ km s$^{-1}$. The colour-coded map (in a logarithmic scale) indicates the number of merging DNSs per cell. The open stars with thin error bars are the masses of the DNSs observed in the Milky Way  with $1-\sigma{}$ errors \citep{martinez2015,antoniadis2016,tauris2017}. The thick error bars show the mass of the NSs in GW170817 within 90\% credible level  assuming the system has low-spin \citep{abbott2017e}.
}
\end{figure}

\begin{figure}
\center{{
\epsfig{figure=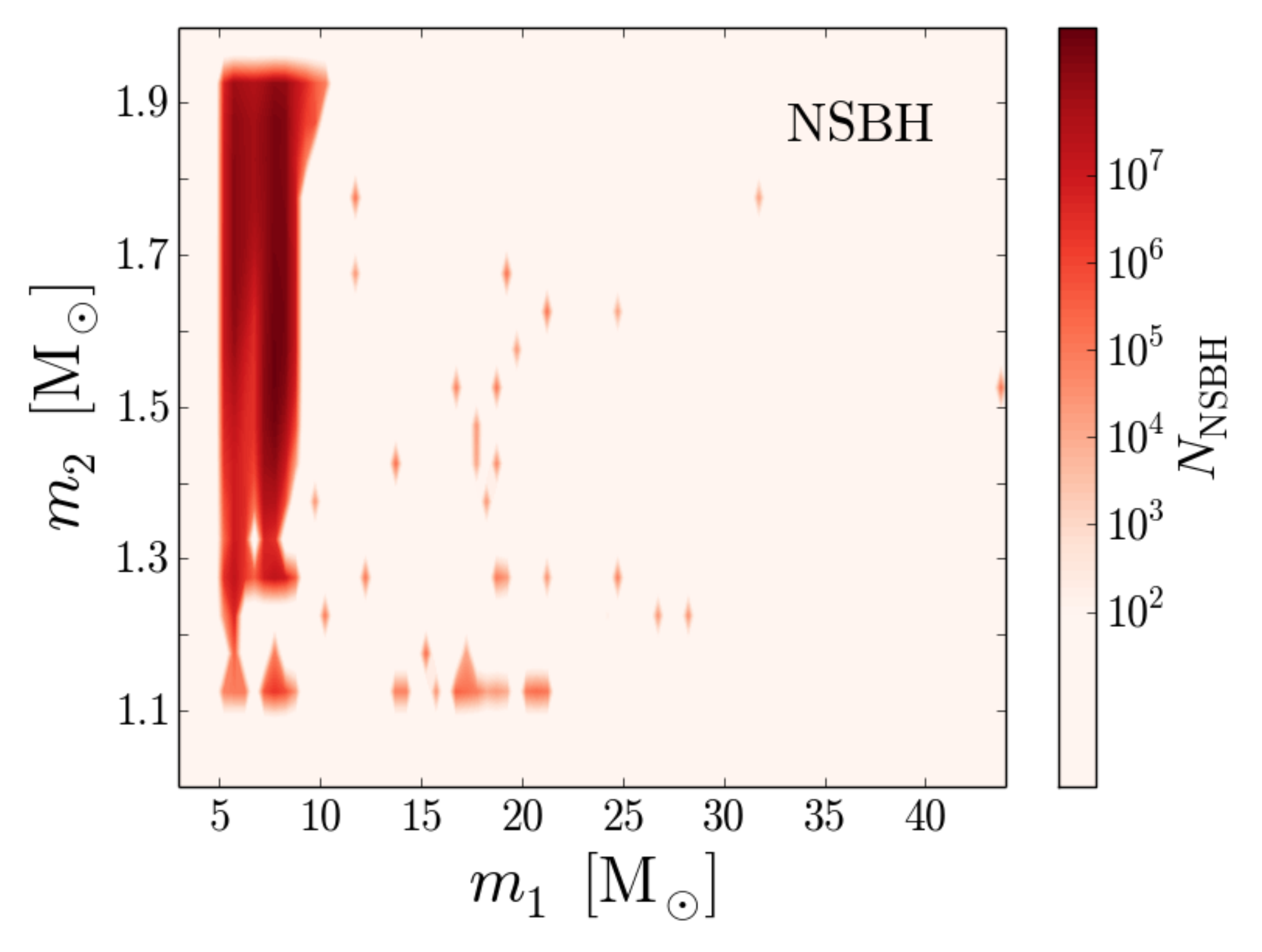,width=9cm} 
}}
\caption{\label{fig:figNSBHmass}
Mass of the NS versus the mass of the BH for all simulated merging NSBHs for the run with $\alpha=5$ and $\sigma{}_{\rm CCSN}=15$ km s$^{-1}$. The colour-coded map (in a logarithmic scale) indicates the number of merging NSBHs per cell. 
}
\end{figure}

\begin{figure}
\center{{
\epsfig{figure=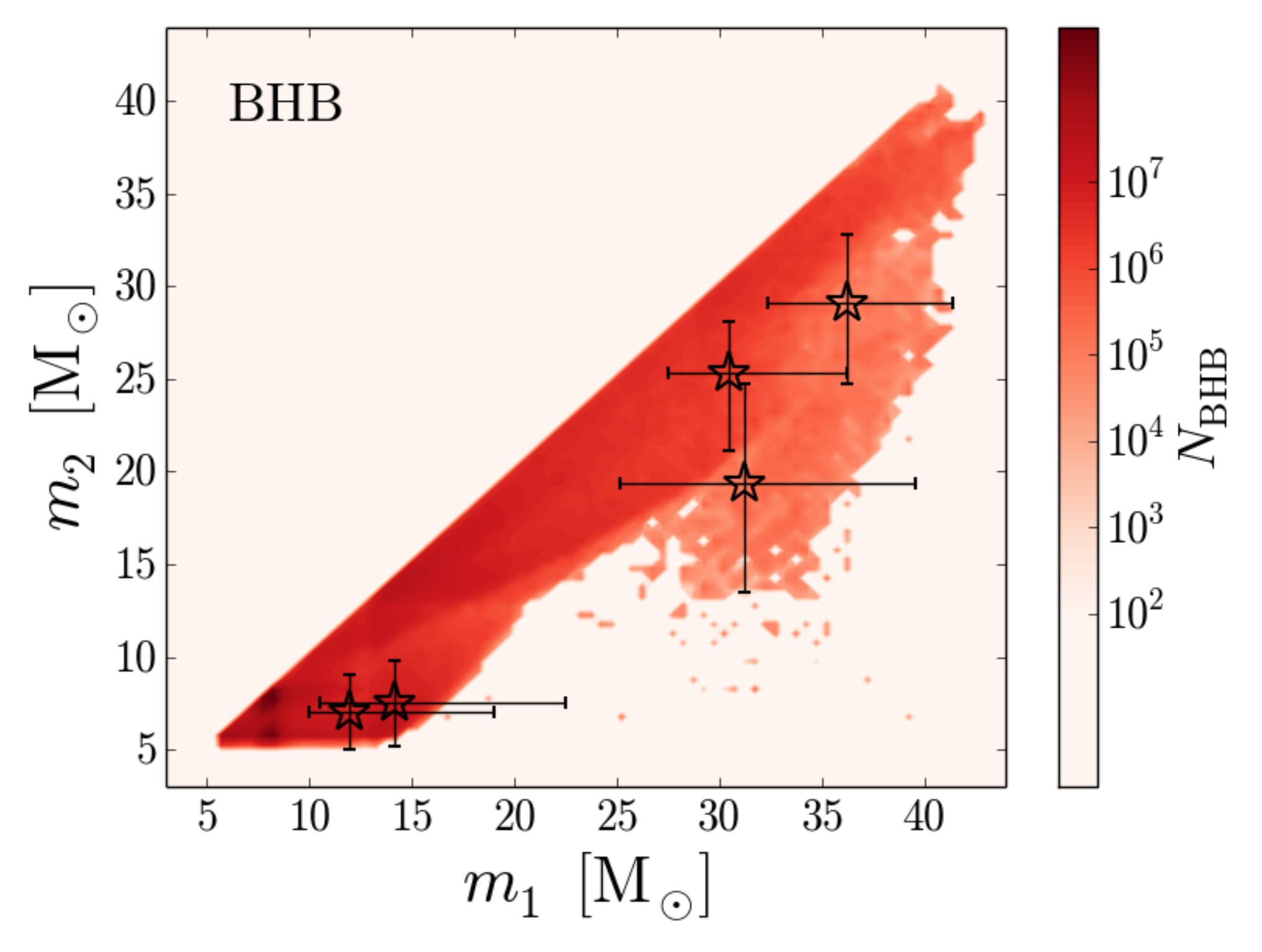,width=9cm} 
}}
\caption{\label{fig:figBHBmass}
Mass of the secondary BH versus the mass of the primary BH for all simulated merging BHBs for the run with $\alpha=5$ and $\sigma{}_{\rm CCSN}=15$ km s$^{-1}$. The colour-coded map (in a logarithmic scale) indicates the number of merging BHBs per cell. The open stars are the five reported GW events associated with BHBs (the error bars show the 90\% credible levels).
}
\end{figure}

From our simulations we can derive the mass distribution of merging DNSs (Fig.~\ref{fig:figNSNSmass}), NSBHs (Fig.~\ref{fig:figNSBHmass}) and BHBs (Fig.~\ref{fig:figBHBmass}) across cosmic time.

Figure~\ref{fig:figNSNSmass} shows the distribution of the mass of the secondary NS (defined as the least massive member of the binary system) with respect to the primary NS (defined as the most massive member of the binary system) for the run with $\alpha=5$ and $\sigma{}_{\rm CCSN}=15$ km s$^{-1}$. Primary masses range from 1.1 to $\sim{}2$ M$_\odot$, the most frequent masses being $\sim{}1.3$ and $\sim{}1.1$ M$_\odot$. Secondary masses also range from 1.1 to $\sim{}1.9$ M$_\odot$, but the vast majority of the secondary NSs has either a low mass ($\sim{}1.1$ M$_\odot$) or a mass very similar to that of the primary. 

In Fig.~\ref{fig:figNSNSmass} we also show the masses of the observed Galactic DNSs and of the components of GW170817. The observed NS masses lie in a region of the $m_1-m_2$ plane populated by a large number of simulated systems ($>10^5$ per grid cell), although the simulated secondary NSs tend to cluster at smaller masses than the observed ones. This might suggest that either the prescriptions of \cite{fryer2012} tend to underestimate NS masses or our population-synthesis simulations underestimate the efficiency of mass accretion onto NSs. We deem the latter hypothesis quite unlikely, because theoretical arguments and observations indicate that the mass accreted by a NS during its life should be $\lesssim{}0.02$ M$_\odot$, unless dynamical interactions allow the NS to exchange stellar companion (see \citealt{tauris2017} and references therein).

Figure~\ref{fig:figNSBHmass} shows the mass of the NS versus the mass of the BH for all merging NSBHs for the run with $\alpha=5$ and $\sigma{}_{\rm CCSN}=15$ km s$^{-1}$. The mass of the NS in these systems spans from $\sim{}$1.1 to $\sim{}2$ M$_\odot$. NS masses $>1.3$ M$_\odot$ are favoured in NSBHs. Most BH masses in merging NSBH systems are $<10$ M$_\odot$ and preferentially $\sim{}5$ M$_\odot$ and $\sim{}8$ M$_\odot$. Few NSBH systems host more massive BHs, up to $\sim{}45$ M$_\odot$. The predominance of small BHs in NSBHs comes from the fact that mass transfer in isolated binaries tends to lead to systems with $m_2/m_1\sim{}1$. 

Finally, Figure~\ref{fig:figBHBmass} shows the mass of the secondary BH (the lighter one) versus the mass of the primary BH (the heavier one) of merging BHBs for the run with $\alpha=5$ and $\sigma{}_{\rm CCSN}=15$ km s$^{-1}$. The mass range spans from 5 to $\sim{}41-43$ M$_\odot$ for both members of the BHB. Light BHBs (with both BH masses $\sim{}5-15$ M$_\odot$) are more common than heavy BHBs (with both BH masses $>15$ M$_\odot$). Figure~\ref{fig:figBHBmass} also shows the mass of the five GW events interpreted as BHB mergers \citep{abbott2016a,abbott2016c,abbott2016d,abbott2017a,abbott2017b,abbott2017c}. All observed merging BHBs lie in regions of the $m_1-m_2$ plane which are densely populated by simulated BHs indicating that our models match the GW observations. Three of the observed GW events (GW150914, GW170104 and GW170814) lie in the region of massive BHBs ($>15$ M$_\odot$), while only two GW events (GW151226 and GW170608) lie in the most densely populated region of low-mass BHBs ($5-15$ M$_\odot$). This is not in tension with our models, because GW interferometers have a higher chance of detecting GWs from massive BHBs than from light BHBs. In a forthcoming study, we will produce mock GW observations of our models, to quantify the agreement between them and GW detections.

\section{Discussion}\label{sec:caveats}
We have shown that it is quite difficult to obtain a DNS merger rate consistent with the one inferred from GW170817: a high value of the CE $\alpha{}$ parameter and low SN kicks are both needed. 

Our models demonstrate that it is possible to predict a merger rate of DNSs, NSBHs and BHBs consistent with those inferred from GW detections with the same assumptions for the CE phase, unlike the conclusions of \cite{chruslinska2018}. The CE is certainly  a process which can make the difference, as already highlighted by \cite{chruslinska2018}. However, we do not need to make any extreme assumptions about the CE phase:  by increasing the value of $\alpha{}$ and by assuming low SN kicks, we recover the merger rate of both DNSs and BHBs. Unlike \cite{chruslinska2018}, we do not need to assume that HG donors survive the CE phase only in the case of NS progenitors, while they merge in the case of BH progenitors. 
The fact that we consider large values of $\alpha{}$ is also one of the most likely reasons why our DNS merger rates are significantly larger than the ones obtained by \cite{kruckow2018}, who assume $\alpha{}<1$ (but our treatment of both binary evolution processes and the cosmological framework are also significantly different from the approach adopted in \citealt{kruckow2018}).


CE  appears once more to be a crucial evolutionary phase for merging compact objects. We stress that our understanding of CE is still far from satisfactory and there is still a large uncertainty on CE (see \citealt{ivanova2013,tauris2017} for a review). 
In this paper, we neglect alternative models for the formation of merging compact objects, which do not require a CE phase. For example, \cite{marchant2016} proposed that two very massive metal-poor stars in a close binary can remain fully mixed because of their tidally induced high spin and thus evolve as a massive over-contact binary: a binary whose components both overfill their Roche lobe but avoid merger, evolving into two nearly equal-mass, very close, Helium stars. These two stars can evolve into a merging BHB or DNS even without CE. According to \cite{mandel2016} and \cite{demink2016}, the two fully-mixed stars can even avoid filling their Roche lobe (because they become Helium stars fast enough), and then evolve into a merging BHB or DNS. These models are not considered in this paper and should be investigated in the future. 


 Our work suggests that low kicks for DNSs (of the order of $\sim{}0-50$ km s$^{-1}$)  are needed to explain the high  merger rate of DNSs inferred from the LIGO-Virgo scientific collaboration.  Other indirect observational estimates of SN kicks give contrasting results (see \citealt{tauris2017} for a recent discussion). \cite{hobbs2005} found that a single Maxwellian with root mean square $\sigma{}_{\rm CCSN}=265$ km s$^{-1}$ can match the proper motions of 233 single pulsars. Other works suggest a bimodal velocity distribution, with a first peak at low velocities (e.g. $\sim{}0$ km s$^{-1}$ according to \citealt{fryer1998} or $\sim{}90$ km s$^{-1}$ according to \citealt{arzoumanian2002}) and a second peak at high velocities ($>600$ km s$^{-1}$ according to \citealt{fryer1998} or $\sim{}500$ km s$^{-1}$ for \citealt{arzoumanian2002}). Similarly, the recent work of \cite{verbunt2017} indicates that a double Maxwellian distribution provides a significantly better fit to the observed velocity distribution than a single Maxwellian. Finally, the analysis of \cite{beniamini2016} shows that low kick velocities ($\lesssim{}30$ km s$^{-1}$) are required to match the majority of Galactic DNSs, especially those with low eccentricity.

Our results are even  more extreme, suggesting that the vast majority of DNSs 
should be born with a relatively low kick velocity.  A possible physical  interpretation for the small natal kicks of DNSs is that these systems undergo ultra-stripped SNe (see \citealt{tauris2017} and references therein for more details). A star can undergo an ultra-stripped SN explosion only if it was heavily stripped by mass transfer to a companion \citep{tauris2013,tauris2015}. The flavour of an ultra-stripped SN can be either an ECSN or an iron core-collapse SN. \cite{tauris2015} suggest that the natal kick of an ultra-stripped SN should be low because of the small mass of the ejecta ($\sim{}0.1$ M$_\odot$). Low kicks ($\lesssim{}50$ km s$^{-1}$) for ultra-stripped iron core-collapse SNe are also confirmed by recent hydrodynamical simulations simulations \citep{suwa2015,janka2017}. Consistently with our results, also \cite{kruckow2018} find significantly larger DNS merger rates when assuming lower kicks for the ultra-stripped SNe (see their table~6).


We have considered only isolated binaries: no dynamical effects are accounted for. Dynamics was demonstrated to have a possibly large impact on the merger rate  of compact objects (see e.g. \citealt{portegieszwart2000,mapelli2013,mapelli2014,ziosi2014,rodriguez2015,rodriguez2016,mapelli2016,askar2016,kimpson2016,banerjee2017,zevin2017,fujii2017}). In particular, single NSs might be efficient in entering DNSs by dynamical exchanges in old globular clusters \citep{belczynski2018}. Alternatively, it is possible that dynamics breaks some BHBs, especially the light ones \citep{zevin2017}, reducing $R_{\rm BHB}$.  The contribution of stellar dynamics might increase the DNS merger rate, relaxing the constraints on CE efficiency and natal kicks we discussed in this paper. We will account for the contribution of dynamics in future works.




Finally, our models suffer from several uncertainties connected with the cosmological framework. In particular, the Illustris simulation predicts a factor of $\sim{}40$ per cent higher SFR at low redshift and a steeper mass-metallicity relation than the observed one. Moreover,  dwarf  galaxies are unresolved in the Illustris. In \cite{mapelli2017} we have estimated that these issues affect the merger rate by a factor of $\sim{}2$. We can compare our results for the DNS merger rate with the complementary approach followed by \cite{giacobbo2018b}. \cite{giacobbo2018b} use the same population-synthesis simulations as we do and derive the local DNS merger rate from the fit to the observed cosmic SFR density \citep{madaudickinson2014}, by making the reasonable assumption that the merger rate of DNSs does not depend on the metallicity of the progenitor stars. For the model with $\alpha=5$ and low SN kicks, \cite{giacobbo2018b} find values of the local merger rate $R_{\rm DNS}\sim{}200-1400$ Gpc$^{-3}$ yr$^{-1}$, which are perfectly consistent with the results presented in this paper ($R_{\rm DNS}\sim{}600$ Gpc$^{-3}$ yr$^{-1}$). Future studies,  adopting other cosmological simulations with different box size and resolution, will help us better quantifying the uncertainty of our predicted rates.



\section{Summary}
We have investigated the merger rate of DNSs, NSBHs and BHBs as a function of redshift, by coupling our population-synthesis simulations with the Illustris cosmological simulation. We have considered three different values of the $\alpha$ parameter describing the efficiency of CE ejection ($\alpha{}=1,3,$ and 5) and we have explored two different cases for the distribution of core-collapse SN kicks (a Maxwellian distribution with 1D rms $\sigma{}_{\rm CCSN}=265$ km s$^{-1}$ and  $\sigma{}_{\rm CCSN}=15$ km s$^{-1}$, respectively).

As already discussed in \cite{mapelli2017}, the merger rate of BHBs steadily increases with redshift reaching a maximum at $z\sim{}2-3$ and then it drops at higher redshift. The merger rate of DNSs and NSBHs follows a similar trend, with a maximum at $z\sim{}1-2$ and at $z\sim{}2-3$ for DNSs and NSBHs, respectively. This trends depends on the cosmic SFR density, on the metallicity evolution  and on the distribution of the delay times between the formation of the progenitor binary system and the merger of the two compact objects.

The BHB merger rate depends only slightly on the value of $\alpha$ and on the 1D rms of the natal kick distribution. In our population-synthesis simulations, we have assumed that the natal kicks of BHs are reduced depending on the amount of fallback; thus natal kicks are low for heavy BHs even when we assume a Maxwellian distribution with high 1D rms ($\sigma{}_{\rm CCSN}=265$ km s$^{-1}$). This explains why the BHB merger rate does not depend significantly on the assumed dispersion of the kick distribution. As already shown in \cite{mapelli2017}, the BHB merger rate strongly depends on the natal kick of BHs if this  is not reduced by the fallback.  

The merger rate of both DNSs and NSBHs strongly depends on the choice of $\alpha{}$ and on the assumed kick distribution. In particular, the merger rate of DNSs increases for a larger value of $\alpha{}$ and  for a lower value of $\sigma{}_{\rm CCSN}$. The merger rate of NSBHs also increases for a lower value of $\sigma{}_{\rm CCSN}$, while the effect of $\alpha$ strongly depends on the choice of $\sigma{}_{\rm CCSN}$.

The current merger rate of BHBs ranges from $\sim{}150$ to $\sim{}240$ Gpc$^{-3}$ yr$^{-1}$, consistent with the merger rate inferred from O1 GW detections although close to the upper limit of the allowed range. The current merger rate of NSBHs spans from $\sim{}10$ (for low natal kicks) to $\sim{}100$ Gpc$^{-3}$ yr$^{-1}$ (for large natal kicks), consistent with the upper limit from the O1 run. Finally, the current merger rate of DNSs ranges from $\sim{}20$ Gpc$^{-3}$ yr$^{-1}$ (for $\alpha=1$ and $\sigma{}_{\rm CCSN}=265$ km s$^{-1}$) to $\sim{}600$ Gpc$^{-3}$ yr$^{-1}$ (for $\alpha=5$ and $\sigma{}_{\rm CCSN}=15$ km s$^{-1}$).

The current merger rate of DNSs is consistent with the one inferred from the detection of GW170817 only if $\alpha\ge{}3$ and only if the natal kick distribution is assumed to be low ($\sigma{}_{\rm ECSN}=\sigma{}_{\rm CCSN}=15$ km s$^{-1}$). A possible physical interpretation is that the vast majority of DNSs form from ultra-stripped SNe, which are expected to produce very small natal kicks ($<50$ km s$^{-1}$, \citealt{suwa2015,tauris2015,tauris2017}).

Finally, we have studied the mass distribution of merging BHBs, NSBHs and DNSs in our simulations. Merging DNS masses range from $\sim{}1.1$ to $\sim{}2$ M$_\odot$ with a prevalence of small masses. Most BHs in merging NSBHs have mass $<10$ M$_\odot$, while the masses of BHs in merging BHBs span from 5 to $\sim{}40$ M$_\odot$. In our simulations we find merging systems with mass consistent with all GW events reported by the LIGO-Virgo collaboration so far.

\section*{Acknowledgments}
We thank the referee, Matthew Benacquista, for his useful comments. We also thank Emanuele Ripamonti, Alessandro Bressan, Elena D'Onghia and Gijs Nelemans for useful discussions. We warmly thank The Illustris team for making their simulations publicly available. Numerical calculations have been performed through a CINECA-INFN agreement and through a CINECA-INAF agreement, providing access to resources on GALILEO and MARCONI at CINECA.  
 MM  acknowledges financial support from the MERAC Foundation through grant `The physics of gas and protoplanetary discs in the Galactic centre', from INAF through PRIN-SKA `Opening a new era in pulsars and compact objects science with MeerKat', from MIUR through Progetto Premiale 'FIGARO' (Fostering Italian Leadership in the Field of Gravitational Wave Astrophysics) and 'MITiC' (MIning The Cosmos  Big Data and Innovative Italian Technology for Frontier Astrophysics and Cosmology), and from the Austrian National Science Foundation through FWF stand-alone grant P31154-N27 `Unraveling merging neutron stars and black hole - neutron star binaries with population-synthesis simulations'. NG acknowledges financial support from Fondazione Ing. Aldo Gini and thanks the Institute for Astrophysics and Particle Physics of the University of Innsbruck for hosting him during the preparation of this paper.
 This work benefited from support by the International Space Science Institute (ISSI), Bern, Switzerland,  through its International Team programme ref. no. 393
 {\it The Evolution of Rich Stellar Populations \& BH Binaries} (2017-18).
 
\bibliography{./bibliography}

\end{document}